\documentclass[aps,prb,twocolumn,groupedaddress]{revtex4}
\usepackage{graphicx}
\usepackage{array}
\input epsf
\usepackage{amsmath,amssymb}

\begin{document}

\title{Inhomogeneous screening of gate electric field by interface states in graphene FETs }
\author{Anil Kumar Singh}
\affiliation{Department of Physics, Indian Institute of Technology Kanpur, Kanpur 208016, India}
\author{Anjan Kumar Gupta}
\affiliation{Department of Physics, Indian Institute of Technology Kanpur, Kanpur 208016, India}
\date{\today}

\begin{abstract}
The electronic states at graphene-SiO$_2$ interface and their inhomogeneity is investigated using the back-gate-voltage dependence of local tunnel spectra acquired with a scanning tunneling microscope. The conductance spectra show two, or occasionally three, minima that evolve along the bias-voltage axis with the back gate voltage. This evolution is modeled using tip-gating and interface states. The energy dependent interface states' density, $D_{it}(E)$, required to model the back-gate evolution of the minima, is found to have significant inhomogeneity in its energy-width. A broad $D_{it}(E)$ leads to an effect similar to a reduction in the Fermi velocity while the narrow $D_{it}(E)$ leads to the pinning of the Fermi energy close to the Dirac point, as observed in some places, due to enhanced screening of the gate electric field by the narrow $D_{it}(E)$.
\end{abstract}
\maketitle

\section{Introduction}
Monolayer graphene is the first experimentally accessible two-dimensional material \cite{Novoselov-2004}, which, together with its linear Dirac-Fermion-like dispersion near Fermi energy, offers access to very exciting physics and applications, such as high speed electronics and photonic devices \cite{Castroneto-2009,Bonaccorso-2010,Sarma-2011,Fiori-2014}. With the objective to investigate the exotic state of electrons in graphene driven by inter-electron interactions, graphene field-effect-transistors (Gr-FETs) of extremely high mobility and free path of tens of microns have been realized \cite{Mayorov-2011,Wang-2013,Banszerus-2016}. However, one still cannot access the Dirac point (DP) in graphene with enough energy resolution due to residual disorder and inhomogeneities \cite{Mayorov-2012}.

In commonly made Gr-FETs with SiO$_2$ gate, the disorder and residual doping are mainly attributed to the interface, defect and/or trap states at graphene-SiO$_2$ interface \cite{Wang-2010,Lee-2011,Xu-2012,Sinha-2014,Lee-2013}. The exact nature of these states depend on the detailed surface structure of amorphous-SiO$_2$ \cite{Nagashio-2011,Zhang-2009,Shi-2009,Fan-2012,Romero-2008,Miwa-2011} together with the species, such as O$_2$ and H$_2$O, adsorbed on it \cite{Xu-2012}. In fact, the nature of doping has been controlled by interface engineering \cite{Wang-2011}. The trapping/detrapping of electrons in these interface states is responsible for noise \cite{Kayyalha-2015} as well as hysteresis in resistance \cite{Wang-2010,Lee-2011,Xu-2012,Sinha-2014,Lee-2013} and capacitance \cite{Kalon-2011,Nagashio-2013} in Gr-FETs. A direct probe of the disorder due to interface states is quantum capacitance \cite{Xia-2009,Ponomarenko-2010,Dröscher-2010,Xu-2011}.

Scanning tunneling microscope (STM) can directly access the electronic states, by local tunneling spectroscopy, which eventually control the electronic properties. Electronic inhomogeneities in graphene as arising from charge disorder due to interface and defect states have been investigated by several STM groups \cite{Deshpande-2009,Deshpande-2011,Samaddar-2016}. Since the carrier density in graphene is small near DP, the tip-gating effect on the spectra is significant \cite{Samaddar-2016,Choudhary-2011}. Additional tip-gating related effects such as ionization of impurities on graphene \cite{Brar-2011}, quantum confinement effects \cite{Zhao-2015}, have also been reported. In addition, as discussed here, the interface states also affect the local tunnel spectra through their weak interaction with graphene.

In this paper, we present a systematic study of local tunnel spectra on several atomically resolved single layer graphene (SLG) surfaces with back-gate voltage ($V_g$). Local tunnel spectra show multiple minima that move along the tip-bias axis as a function of $V_g$. This evolution of the minima is modeled using tip-gating and an energy dependent interface states' density. The later is found to be spatially inhomogeneous with a narrow energy-width in some places pinning the graphene Fermi energy and broad width in other places leading to an apparent reduction in Fermi velocity. Finally, we discuss the possible origin and implications of these inhomogeneous interface states.

\section{Experimental details}
\begin{figure}
\includegraphics[width=8.5cm]{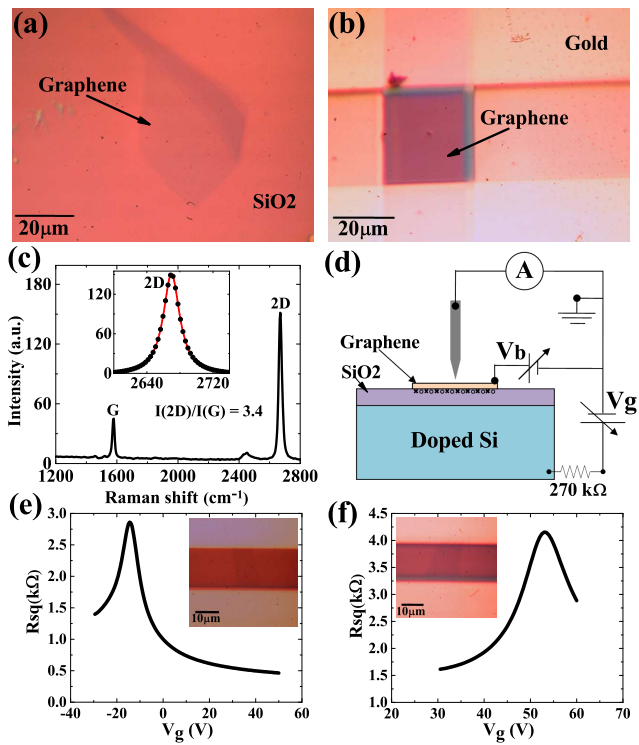}
\caption{(a) Optical image of graphene on SiO$_2$. (b) shows the same graphene flake after deposition of Cr(10nm)/Au(50nm) contact. (c) Raman Spectra of a graphene flake using 532nm and 0.50 mW laser; the inset shows enlarged view of the 2D peak with Lorentzian fitting. (d) shows the electrical schematic of the STM measurement. (e) and (f) show the two probe resistance as a function of gate voltage on piranha and oxygen-plasma treated substrates, respectively.}
\label{fig:device}
\end{figure}

Graphene was mechanically exfoliated from Kish graphite using adhesive tape on n-doped Si substrate with 300 nm thick thermal oxide. The substrate was first cleaned either with oxygen plasma (50 W) for 5 minute or by dipping into the freshly prepared piranha solution for 5 min followed by rinsing in de-ionized water and blow drying. Exfoliation was done within 30 min of the cleaning process. The Raman spectrum of a sample, in Fig.\ref{fig:device}(c), shows characteristic Raman features, i.e. G and 2D bands, of SLG. The absence of D-peak indicates lack of defects. Single Lorentzian fitting of the 2D peak [see inset Fig.\ref{fig:device} (c)] with I(2D)/I(G)= 3.4 confirms SLG \cite{Ferrari-2006}. Initially, to avoid contamination from wet chemical process and resist used in lithography steps, a mechanical shadow-masking method \cite{Bao-2010} was followed for electrical contacts. Cr(10nm)/Au(50nm) was deposited twice after masking the graphene flake using 50$\mu$m (or 25$\mu$m) diameter tungsten wires [see Fig.\ref{fig:device}(b)] under optical microscope. The wires were kept perpendicular to each other in the two steps for ease of aligning graphene with STM tip under optical microscope. We note that the metal induced doping for Cr/graphene contacts is negligible \cite{Nagashio-2010}. Later on we found that electron beam lithography with Cr/Au lift-off followed by Ar/H$_2$(5\% H$_2$ in Argon) annealing for 2 hrs at 350$^\circ$C also leads to similar atomic resolution images and spectra.

Several freshly prepared as well as stored, in vacuum desiccators over several months time period, SLG samples were studied using a homemade vacuum-STM at room-temperature with an integrated 2D-positioner \cite{Gupta-2008} for coarse sample movement. The STM tip was aligned with the gold pad near graphene in ambient conditions followed by transfer of STM to the vacuum chamber. The chamber was then pumped to a pressure lower than 5 x $10^{-4}$ mbar using a cryopump attached to the chamber. Guided by the STM images, the sample was coarse-adjusted in-situ to align graphene in front of the STM tip and also to explore larger area. As shown in Fig.\ref{fig:device}(d), $V_g$ was applied to the silicone substrate with 270 k$\Omega$ series resistance. The STM bias voltage ($V_b$) was applied on graphene while the tip was kept at (virtual) ground potential. Electrochemically etched and hydrofluoric acid treated tungsten wire (0.25 mm diameter) was used as the STM tip. The apex radius of the tip was found to be in 30-50 nm range from electron microscopy. The tunnel conductance spectra were acquired by using 20mV amplitude ac-modulation with the dc bias voltage. The general reproducibility of the spectra was confirmed on different regions of several samples with a number of tungsten tips.

\section{Experimental results}

We performed STM/S on twelve different SLG devices and transport on a few devices prepared by either piranha or oxygen-plasma cleaning. The sheet resistance of a device made using piranha cleaning, in Fig.\ref{fig:device}(e), shows the expected \cite{Venugopal-2012} n-type doping with DP occurring at V$_g$=-14V. SLG device made by oxygen plasma cleaning, in Fig.\ref{fig:device}(f), shows large p-type doping with DP occurring at V$_g$=53V. Oxygen-plasma treatment increases the silanol group density by removing hydrocarbon contaminants on SiO$_2$ and hence p-dopes graphene \cite{Nagashio-2011}.

\begin{figure}
\includegraphics[width=8.5cm]{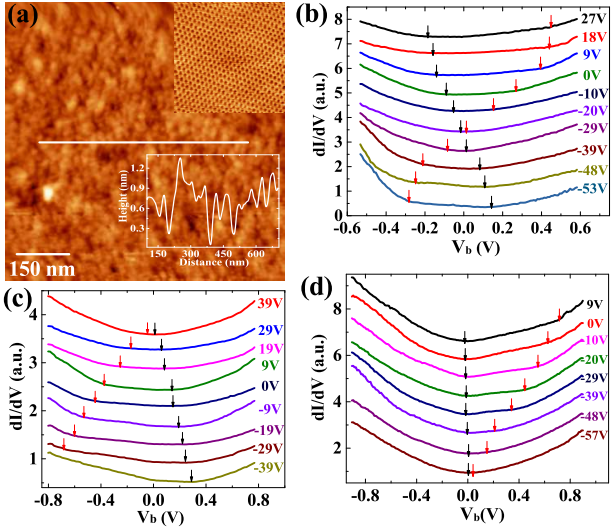}
\caption{ (a) STM image of SLG (0.6 V/0.1 nA, 0.8$\times$0.8 $\mu$m$^2$) with a line-cut along the marked line. The inset shows a zoomed-in image (0.4 V/0.1 nA, 7.9$\times$7.9 nm$^2$) showing honeycomb graphene lattice. (b), (c) and (d) show V$_g$ variation of the local tunnel spectra on three different SLG devices having moderate n-, large p-, and large n-type doping, respectively. All the spectra were taken with same set-point current of 0.1nA but at V$_b$ = 0.6, 0.8, 0.8V, respectively. The black and red arrows mark the location of primary and secondary minima, respectively.}
\label{fig:result}
\end{figure}

The STM/S was done at lateral distances of more than 1$\mu$m from the metal-graphene interface so as to avoid the influence of the metal contact. Large area topographic image of SLG in Fig.\ref{fig:result}(a) shows 0.4 nm rms roughness due to underlying SiO$_2$ \cite{Deshpande-2009,Choudhary-2011}. The inset shows a zoomed-in image showing atomically resolved surface with honeycomb structure. Fig.\ref{fig:result}(b), (c) and (d) show the evolution of the tunnel spectra with V$_g$ acquired on three different samples with different dopings. Fig.\ref{fig:result}(b) shows two minima moving in opposite directions along the bias axis with V$_g$. One of the minima, called primary minima, occurs at $V_b=V_{b}^{1}$ when the tip's Fermi energy coincides with the DP energy of graphene and the other one, called secondary minimum, occurs at $V_b=V_{b}^{2}$ when the Fermi energy of graphene nearly coincides with its DP energy \cite{Choudhary-2011}. Fig.\ref{fig:result}(c) shows tunnel spectra on a SLG device with large p-doping, where the two minima move towards each other when V$_g$ is increased from -39V and the two eventually meet near V$_g$ = 40V. The tunnel spectra in Fig.\ref{fig:result}(d) also show two minima but only $V_{b}^{2}$ changes with V$_g$ while $V_{b}^{1}$ remains fixed near $V_{b}=0$ for V$_g$ values from -53 to 46V. From our study on these SLG devices we find that the tunnel spectra on devices prepared by piranha cleaning show mostly n-type doping (i.e. $V_{b}^{2}$ at positive bias voltage for V$_g$ = 0) consistent with their transport behavior while those on oxygen plasma processed devices show mostly p-type doping, again consistent with the respective transport behavior. The evolution of the two minima with V$_g$ in local spectra shows significant variation even on a given SLG device, which cannot be modeled by simple tip-gating effect \cite{Choudhary-2011} as discussed further.

V$_g$ dependence of the local tunnel-spectra has been studied and modeled by several STM groups [\onlinecite{Samaddar-2016,Choudhary-2011,Zhao-2015}] using tip-gating effect. In this model the primary minima position shows $v_F\sqrt{|V_g-V_{g}^{D}|}$ dependence on V$_g$ with $v_F$ as the Fermi velocity of graphene and $V_{g}^{D}$ is a local constant dependent on local doping as described later. In order to fit the V$_g$-dependence of the two minima as arising from tip-gating effect we will need $ 1.0 \times 10^5 m/s \leqslant v_F \leqslant 7.5 \times 10^5 m/s$. From the Friedel oscillations, seen using STM, near atomic defects in Ar$^+$ ion-irradiated graphene, Tapasztó et al. \cite{Tapasztó-2008} found three times reduction in $v_F$  and attributed it to induced disorder in hopping amplitudes. However, we cannot understand the origin of this ten times reduction in v$_F$ as our exfoliated graphene is unlikely to have such atomic defects. Such hard defects should also give rise to the defect mediated D-peak in the Raman spectra which we do not see at all, see Fig.\ref{fig:device}(c). Another drawback of v$_F$ reduction in the simple tip-gating model is the disappearance of secondary minima. The later can be recovered but only with significant reduction in tip-sample distance, which again cannot be justified. Thus we conclude that the simple tip-gating model cannot explain various tunnel spectra observed in our experiments. On the other hand, interface states have been invoked for understanding various experimental results as discussed earlier, which we incorporate in the simple tip-gating model to understand our local spectra as follows.

\section{Effect of interface states on tunnel spectra}

As discussed earlier, the interface states between graphene and SiO$_2$ can arise from the detailed SiO$_2$ surface structure and adsorbates. The STM tip has two interactions with graphene, namely, electrostatic (tip-gating) and tunneling. The former directly affects the filling of the graphene states and the interface states, which weakly interact with graphene, while the electron tunneling happens only between graphene and the tip states. The interface states will not directly affect the tunnel conductance if the equilibration rate of the tunneling electrons is much larger than the electron transfer rate between graphene and the interface states. The electron equilibration time within graphene is expected to be sub-ns while typical electron transfer times for the later process are more than $\mu$s order. The other measurement times in the STM/S are much larger. Thus in this quasi-static equilibrium limit the occupancy of the interface and graphene states will be governed by the same Fermi distribution function. Here, we discuss a model that incorporates the effect of energy dependent interface-state density $D_{it}(E)$ (defined as the number of states per unit area and per unit energy) and tip-gating to find V$_b$ and V$_g$ dependent tunnel conductance.

\begin{figure}
\includegraphics[width=9.0cm]{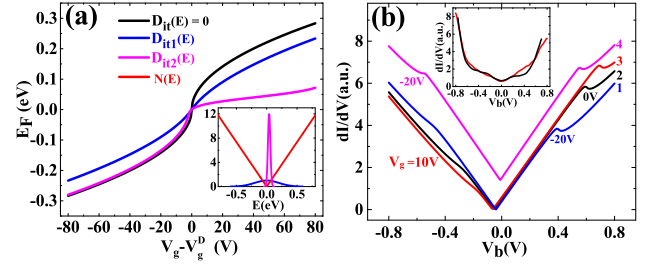}
\caption{(a) shows the effect of $D_{it}(E)$ on movement of $E_F$ as a function of $({V_{g}-V_{g}^{D}})$ with the inset showing $D_{it1}$, $D_{it2}$ and graphene density of states (N(E)) in 10$^{13}$cm$^{-2}$-eV$^{-1}$ units. (b) shows the calculated tunnel conductance for gaussian $D_{it2}(E)$ (curve-1, 2 and 3); curve-4 corresponds to another $D_{it}(E)$ discussed in the text. The inset in (b) shows experimental spectra with three minima captured by curve-4.}
\label{fig:model-plots}
\end{figure}

The density of states of SLG near the DP is given by,
\begin{equation}
\small N(E)=2|{E+E_{F}}|/\pi{(\hbar v_F)^{2}}\\
\label{eq:DOS}
\end{equation}
with $E_{F}$ as the Fermi energy of graphene measured from the DP. A change in $V_g$ by $dV_g$ causes a change in $E_F$ by $dE_F$ and a change in total charge density by $d\sigma$. The latter change occurs due to change in filling of the graphene, as well as the interface trap, states. $dV_g$ is shared between $dE_F$ and the potential drop across the gate oxide giving $dV_{g}=-({d\sigma}/{C_{ox}})+({dE_F}/{e})$. Here $e$ is the magnitude of electronic charge and $C_{ox}= {\kappa\epsilon_{0}}/{d_{ox}}$ with $\kappa\approx 4$ as the dielectric constant of SiO$_2$, $\epsilon_{0}$ as the free space permeability and  $d_{ox}$ as the SiO$_2$ thickness. With $d\sigma$, due to graphene and interface states, as $d\sigma=-e[N(E_F) + D_{it}(E_F)]dE_F$ we get $dV_{g}=({e}/{C_{ox}})[N(E_F) + D_{it}(E_F)]dE_F + ({dE_F}/{e})$. On integration this gives, $e(V_{g}-V_{g}^{D})= \text{Sgn}(E_{F})[{e^2 E_F^{2}}/{C_{ox} \pi(\hbar v_F)^{2}}] + ({e^2}/ {C_{ox}})\int_{0}^{E_{F}} D_{it}(E)dE + E_{F}$. Here, $V_{g}^{D}$ as a local constant, whose value depends on various contact potentials and the interface trap density. It can be found by $V_g$ required to make $E_F$ coincide with the DP. In addition, incorporating the effect of tip-gating [\onlinecite{Choudhary-2011}], we get,
\begin{equation}
\small e(V_{g}-V_{g}^{D}-\beta V_{b})= \frac {\text{Sgn}(E_{F})e^2 E_F^{2}} {C_{ox} \pi(\hbar v_{F})^{2}} + E_{F} +
\frac {e^2} {C_{ox}}\int_{0}^{E_{F}}{D_{it}(E)}dE.\\
\label{eq:EF}
\end{equation}
Here, $\beta={d_{ox}}/{z\kappa}$ with $z$ as the tip-sample separation. Right hand side of Eq.\ref{eq:EF} is a monotonically increasing single valued function of $E_{F}$. Thus the tunnel conductance at given $V_b$ and $V_g$ and at zero temperature will be given by $G(V_b,V_g)\propto|{eV_{b} +E_{F}(V_g,V_b)}|$.

We take an energy-localized $D_{it}(E)$ with Gaussian shape, i.e.
\begin{equation}
\begin{aligned}
D_{it}(E)= ({D_{it0}}/{\sqrt{2\pi}\delta})\text{exp}(-[({E-\epsilon_{c}})/{\sqrt{2}\delta}]^{2}).
\end{aligned}
\label{eq:distribution}
\end{equation}
Here $D_{it0}$ represents total defect state density, $\epsilon_c$ is the center energy and $\delta$ represents the energy-width. In this case Eq.\ref{eq:EF} gives,
\begin{equation}
\begin{aligned}
e(V_{g}&-V_{g}^{D} -\beta V_{b})= \text{Sgn}[E_{F}](E_F^{2}/{e\gamma^2}) + E_{F} +\\ &
{e^2 D_{it0} d_{ox}}/{2\kappa\epsilon_{0}}[\text{erf}({\epsilon_{c}}/{\sqrt{2}\delta}) + \text{erf} (({E_{F}-\epsilon_{c}})/ {\sqrt{2}\delta})]
\end{aligned}
\label{eq:gaussian}
\end{equation}
Fig.\ref{fig:model-plots}(a) shows $E_{F}$ as a function of $V_{g}-V_{g}^{D}$ for two different $D_{it}(E)$ (shown in inset). $D_{it1}(E)$ is broad in energy and $D_{it2}(E)$ is narrow. For $D_{it1}(E)$, dependence of $E_{F}$ on ${V_{g}-V_{g}^{D}}$ shows qualitatively similar behavior as $D_{it}(E)=0$ except the reduced movement of the primary minima with V$_g$. The narrow $D_{it2}(E)$ shows pinning of $E_{F}$. The $V_g$-range over which $E_{F}$ remains pinned depends on $D_{it0}$ and $\delta$ while pinning-energy depends on $\epsilon_{c}$. Curves 1, 2 and 3 of Fig.\ref{fig:model-plots}(b) show the calculated tunnel conductance for $D_{it2}(E)$ with $V_{g}^{D}=-58V$ and z$=0.75$nm, at different $V_g$ voltages. Here $V_{b}^{2}$ moves with $V_g$ while $V_{b}^{1}$ remains pinned near $V_{b}=0$. At negative $V_b$ there is a feature showing the possibility of another minima and moving away from $V_{b}^{1}$ when $V_g$ changes from 10V to -20V. Similar features and movement with $V_g$ have been seen in experiments, see Fig.\ref{fig:result}(d). The curve-4 in Fig.\ref{fig:model-plots}(b) shows calculated tunnel conductance with three minima for another $D_{it}(E)$ with $D_{it0}=6\times10^{12}/cm^2$, $\epsilon_{c}=0.01$eV and $\delta=0.01$eV. The inset shows a few experimental spectra with three minima.
\begin{figure}
\includegraphics[width=8.5cm]{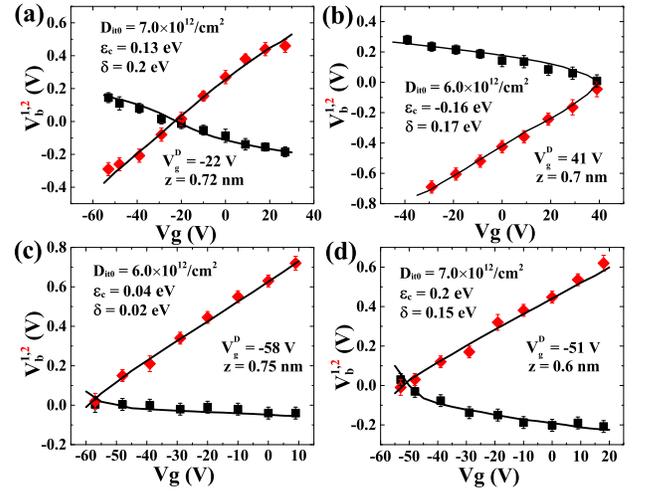}
\caption{Variation of the $V_{b}^{1}$ (squares) and $V_{b}^{2}$ (rhombus) with $V_{g}$ with the solid lines showing the calculated position of the two minima. The plots in (c) and (d) are from two locations of the same sample while those in (a) and (b) are from two other samples. The plot in (b) is from O$_2$-plasma cleaned sample, while the other three are from piranha cleaned ones. The plots in (a), (b) and (c) are derived from (b), (c) and (d), respectively, of fig.\ref{fig:result}. The fitting parameters used in $D_{it}(E)$ (see Eq.\ref{eq:distribution}) are shown in respective plots.}
\label{fig:summary}
\end{figure}

\section{Analysis and discussion}
Fig.\ref{fig:summary} shows the evolution of $V_{b}^{1}$ and $V_{b}^{2}$ of experimentally observed tunnel spectra, on three representative devices, together with the fitting. The plots of Fig.\ref{fig:summary}(a), (b) and (c) correspond to those in Fig.\ref{fig:result}(b), (c) and (d), respectively. Fig.\ref{fig:summary}(c) and (d) are from spectra taken at two different locations of the same sample. We note that the slope of $V_{b}^{2}$ Vs $V_g$ line does not change much between spectra while the $V_g$-dependence of $V_{b}^{1}$ has significant variation. The former depends on $\beta={d_{ox}}/{z\kappa}$, which does not change significantly between spectra while the later depend on $\delta$, which has significant variation. Small $\delta$ leads to a pinned $E_F$ and a large $\delta$ only slows down the $V_g$ dependence of $E_F$. Inhomogeneity in $\delta$, see [Fig.\ref{fig:summary}(c) and (d)], for a given sample is common to all our studied devices and indicates an inhomogeneous screening of the gate electric field by the interface states.

The DFT as well as ab initio calculations show broad or a constant $D_{it}(E)$ arising from silanol group on the SiO$_2$ surface and narrow $D_{it}(E)$ just above DP as arising from a partially occupied state below the conduction band minimum of SiO$_2$ \cite{Romero-2008,Miwa-2011}. The later can donate electrons and pin $E_F$ above DP. The narrow states can come from the formation of three-fold coordinated O-atoms in amorphous-SiO$_2$ \cite{Miwa-2011} due to higher local concentration of Si-atoms or nearby O-vacancies. The energy-localized trap/interface states with a Gaussian distribution have also been invoked recently \cite{Terrs-2016} to understand the transport in graphene constrictions.

\section{Conclusions}
In conclusion our STM/S study on several atomically resolved SLG surfaces with back-gate show multiple minima in local spectra that move along the tip-bias axis as a function of $V_g$. The evolution of the minima is successfully modeled using tip-gating and an energy dependent interface states' density. The later is found to be inhomogeneous and even leads to a pinning of the graphene Fermi energy in some places. Such inhomogeneous screening of the gate electric field by the interface states will lead to a non-linear change in the carrier density with gate-voltage together with a mobility change due to the change in potential-landscape seen by graphene due to the change in filling, with $V_g$, of the interface/trap states.
\section{Acknowledgement}
A.K.G. and A.K.S. acknowledge financial support from DST and MHRD of the Government of India, respectively.


\begin{thebibliography}:

\bibitem{Novoselov-2004}  K. S. Novoselov, A. K. Geim, S. V. Morozov, D. Jiang, Y. zhang, S. V. Dubonos, I. V. Grigorieva and A. A. Firsov, Science {\bf 306}, 666 (2004).
\bibitem{Castroneto-2009}  A. H. Castro Neto, F. Guinea, N. M. R. Peres, K. S. Novoselov and A. K. Geim, Rev. Mod . Phys. {\bf 81}, 109 (2009).
\bibitem{Bonaccorso-2010}  F. Bonaccorso, Z. Sun, T. Hasan and A. C. Ferrari, Nat. Photonics {\bf 4}, 611 (2010).
\bibitem{Sarma-2011}  S. Das Sarma, Shaffique Adam, E. H. Hwang and Enrico Rossi, Rev. Mod . Phys. {\bf 83}, 407 (2011).
\bibitem{Fiori-2014}  G. Fiori, F. Bonaccorso, G. Iannaccone, T. Palacios, D. Neumaier, A. Seabaugh, S. K. Banerjee and L. Colombo, Nat. Nanotech. {\bf 9}, 768 (2014).
\bibitem{Mayorov-2011}  A. S. Mayorov, R. V. Gorbachev, S. V. Morozov, L. Britnell, R. Jalil, L. A. Ponomarenko, P. Blake, K. S. Novoselov, K. Watanabe, T. Taniguchi and A. K. Geim, Nano Lett. {\bf 11}, 2396 (2011).
\bibitem{Wang-2013} L. Wang, I. Meric, P. Y. Huang, Q. Gao, Y. Gao, H. Tran, T. Taniguchi, K. Watanabe, L. M. Campos, D. A. Muller, J. Guo, P. Kim, J. Hone, K. L. Shepard and C. R. Dean, Science {\bf 342}, 614 (2013).
\bibitem{Banszerus-2016}  L. Banszerus, M. Schmitz, S. Engels, M. Goldsche, K. Watanabe, T. Taniguchi, B. Beschoten and C. Stampfer, Nano Lett. {\bf 16}, 1387 (2016).
\bibitem{Mayorov-2012}  A. S. Mayorov, D. C. Elias, I. S. Mukhin, S. V. Morozov, L. A. Ponomarenko, K. S. Novoselov, A. K. Geim and R. V. Gorbachev, Nano Lett. {\bf 12}, 4629 (2012).
\bibitem{Wang-2010} H. Wang, Y. Wu, C. Cong, J. Shang and T. Yu, ACS Nano {\bf 4}, 7221 (2010).
\bibitem{Lee-2011} Y. G. Lee, C. G. Kang, U. J. Jung, J. J. Kim, H. J. Hwang, H. J. Chung, S. Seo, R. Choi and B. H. Lee, Appl. Phys. Lett. {\bf 98}, 183508 (2011).
\bibitem{Xu-2012} H. Xu, Y. Chen, J. Zhang and H. Zhang, small {\bf 8}, 2833 (2012).
\bibitem{Sinha-2014} D. Sinha and J. U. Lee , J. Appl. Phys. {\bf 116}, 074516 (2014).
\bibitem{Lee-2013} Y. G. Lee, C. G. Kang, C. Cho, Y. Kim, H. J. Hwang and B. H. Lee, Carbon {\bf 60}, 453 (2013).
\bibitem{Nagashio-2011} K. Nagashio,T. Yamashita, T. Nishimura, K. Kita and A. Toriumi, J. Appl. Phys. {\bf 110}, 024513 (2011).
\bibitem{Zhang-2009} Y. Zhang, V. W. Brar, C. Girit, Y. Yayon, A. Zettl and M. F. Crommie, Nat. Phys. {\bf 5}, 722 (2009).
\bibitem{Shi-2009} Y. Shi, X. Dong, P. Chen, J. Wang and L. J. Li, Phys. Rev. B {\bf 79}, 115402 (2009).
\bibitem{Fan-2012} X. F. Fan, W. T. Zheng, V. Chihaia, Z. X. Shen and J. L. Kuo, J. Phys. Condens. Matter {\bf 24}, 305004 (2012).
\bibitem{Romero-2008} H. E. Romero, N. Shen, P. Joshi, H. R. Gutierrez, S. A. Tadigadapa, J. O. Sofo and P. C. Eklund, ACS Nano {\bf 2}, 2037 (2008).
\bibitem{Miwa-2011} R. H. Miwa, T. M. Schmidt, W. L. Scopel and A. Fazzio, Appl. Phys. Lett. {\bf 99}, 163108 (2011).
\bibitem{Wang-2011} R. Wang, S. Wang, D. Zhang, Z. Li, Y. Fang and X. Qiu, ACS Nano {\bf 5}, 408 (2011).
\bibitem{Kayyalha-2015} M. Kayyalha and Y. P. Chen , Appl. Phys. Lett. {\bf 107}, 113101 (2015).
\bibitem{Kalon-2011} G. Kalon, Y. J. Shin, V. G. Truong, A. Kalitsov and H. Yang, Appl. Phys. Lett. {\bf 99} 083109 (2011).
\bibitem{Nagashio-2013} K. Nagashio,T. Nishimura and A. Toriumi, Appl. Phys. Lett. {\bf 102}, 173507 (2013).
\bibitem{Xia-2009}  J. Xia, F. Chen, J. Li and Nongjian Tao, Nat. Nanotech. {\bf 4}, 505 (2009).
\bibitem{Ponomarenko-2010}  L. A. Ponomarenko, R. Yang, R. V. Gorbachev, P. Blake, A. S. Mayorov,K. S. Novoselov, M. I. Katsnelson and A. K. Geim, Phys. Rev. Lett. {\bf 105}, 1368019 (2010).
\bibitem{Dröscher-2010} S. Droscher, P. Roulleau, F. Molitor, P. Studerus, C. Stampfer, K. Ensslin, and T. Ihn, Appl. Phys. Lett. {\bf 96}, 152104 (2010).
\bibitem{Xu-2011} H. Xu, Z. Zhang, and L. M. Peng, Appl. Phys. Lett. {\bf 98}, 133122 (2011).
\bibitem{Deshpande-2009} A. Deshpande, W. Bao, F. Miao, C. N. Lau and B. J. LeRoy, Phys. Rev. B {\bf 79}, 205411 (2009).
\bibitem{Deshpande-2011} A. Deshpande, W. Bao, F. Miao, C. N. Lau and B. J. LeRoy, Phys. Rev. B {\bf 83}, 155409 (2011).
\bibitem{Samaddar-2016} S. Samaddar, I. Yudhistira, S. Adam, H. Courtois and C. B. Winkelmann, Phys. Rev. Lett. {\bf 116}, 126804 (2016).
\bibitem{Choudhary-2011} S. K. Choudhary and A. K. Gupta, Appl. Phys. Lett. {\bf 98}, 102109 (2011).
\bibitem{Brar-2011} V. W. Brar, R. Decker, H. M. Solowan, Y. Wang, L. Maserati, K. T. Chan, H. Lee, C. O. Girit, A. Zettl, S. G. Louie, M. L. Cohen and M. F. Crommie, Nat. Phys. {\bf 7}, 43 (2011).
\bibitem{Zhao-2015} Y Zhao, J. Wyrick, F. D. Natterer, J. F. Rodriguez-Nieva, C. Lewandowski, K. Watanabe,T. Taniguchi, L. S. Levitov, N. B. Zhitenev and J. A. Stroscio, Science {\bf 348}, 672 (2015).
\bibitem{Ferrari-2006} A. C. Ferrari, J. C. Meyer, V. Scardaci, C. Casiraghi, M. Lazzeri, F. Mauri, S. Piscanec, D. Jiang, K. S. Novoselov, S. Roth and A. K. Geim, Phys. Rev. Lett. {\bf 97}, 187401 (2006).
\bibitem{Bao-2010} W. Bao, G. Liu, Z. Zhao, H. Zhang, D. Yan, A. Deshpande, B. J. LeRoy and C. N. Lau, Nano Res. {\bf 3}, 98 (2010).
\bibitem{Nagashio-2010} K. Nagashio, T. Nishimura, K. Kita and A. Toriumi, Japanese J. Appl. Phys. {\bf 49}, 051304 (2010).
\bibitem{Gupta-2008} A. K. Gupta, R. S. Sinha, and R. K. Singh, Rev. Sci. Instrum. {\bf 79}, 063701 (2008).
\bibitem{Venugopal-2012} Archana Venugopal, Doctor of Philosophy in electrical engineering, The university of texas at dallas, 2012.
\bibitem{Tapasztó-2008} L. Tapaszto, G. Dobrik, P. Nemes-Incze, G. Vertesy, Ph. Lambin and L. P. Biró, Phys. Rev. B {\bf 78}, 233407 (2008).
\bibitem{Terrs-2016} B. Terres, L. A. Chizhova, F. Libisch, J. Peiro, D. Jorger, S. Engels, A. Girschik, K. Watanabe, T. Taniguchi, S.V. Rotkin, J. Burgdorfer and C. Stampfer, Nat Commun. {\bf 7}, 11528 (2016).

\end{thebibliography}
\end{document}